\documentclass{webofc}
\usepackage[varg]{txfonts}   
\usepackage{subfig}
\usepackage{comment}
\begin{document}
\title{Predicting nucleon-nucleus scattering observables using nuclear structure theory}
%
%

\author{\firstname{Aaina} \lastname{Thapa}\inst{1}\fnsep\thanks{\email{aaina1@llnl.gov}} \and
        \firstname{Jutta} \lastname{Escher}\inst{1} \and
        \firstname{Emanuel} \lastname{Chimanski}\inst{2} \and
         \firstname{Marc} \lastname{Dupuis}\inst{3,4} \and
         \firstname{Sophie} \lastname{Péru}\inst{3,4}\and
         \firstname{Walid} \lastname{Younes}\inst{1}
}

\institute{Lawrence Livermore National Laboratory, Livermore, CA 
\and
          Brookhaven National Laboratory, Upton, NY 
\and
           CEA, DAM, DIF, F-91297 Arpajon, France and Université Paris-Saclay, CEA 
\and
           Laboratoire Matière sous Conditions Extrêmes, 91680 Bruyères-Le-Châtel, France
          }

\abstract{%
Developing a predictive capability for inelastic scattering will find applications in multiple areas. Experimental data for neutron-nucleus inelastic scattering is limited and thus one needs a robust theoretical framework to complement it. Charged-particle inelastic scattering can be used as a surrogate for $(n, \gamma)$ reactions to predict capture cross sections for unstable nuclei. Our work uses microscopic nuclear structure calculations for spherical nuclei to obtain nucleon-nucleus scattering potentials and calculate cross sections for these processes. We implement the Jeukenne, Lejeune, Mahaux (JLM) semi-microscopic folding approach, where the medium effects on nuclear interaction are parameterized in nuclear matter to obtain the nucleon-nucleon $(NN)$ interaction in a medium at positive energies. We solve for the nuclear ground state using the Hartree-Fock-Bogliubov (HFB) many-body method, assuming the nucleons within the nucleus interact via the Gogny-D1M potential. The vibrational excited states of the target nucleus are calculated using the quasi-particle random phase approximation (QRPA). We demonstrate our approach for spherical nuclei in the medium-mass region, showing scattering results for the $^{90}$Zr nucleus. }

\maketitle
\section{Introduction}
\label{intro}
The broader goal of this work is to develop a predictive microscopic approach for inelastic scattering. Inelastic scattering experiments have long been used to study collective excitations in stable nuclei. The idea is to build and implement a framework which can be used, for stable and unstable nuclei alike, by combining nuclear structure theory with existing reaction codes. Our approach combines state-of-the-art HFB and QRPA nuclear structure methods with scattering calculations. HFB and QRPA many-body methods can be used for all nuclei across the nuclear chart. We use the JLM single-folding approach to take results from structure calculations as input and generate scattering potentials. Our implementation of folding is designed such that it can be extended to use results from other many-body methods, like the nuclear shell model or $ab$-$initio$ approaches, and other local effective $NN$ interactions at positive energies. It is being developed with the intent to predict the effect of the microscopic structure of the target on direct inelastic scattering systematically for all isotopes, going from proton to neutron drip lines for any element of interest. This can support efforts of radioactive beam facilities in measuring collective excitations for unstable nuclei. The applications of our developments include enabling indirect ('surrogate') experimental constraints for calculating neutron-capture cross sections and improving neutron-nucleus data evaluations. \\

Charged-particle inelastic scattering can be used as a surrogate reaction mechanism~\cite{Ratkiewicz:19prl,Escher:18prl,Escher:16a,Escher:12rmp,PerezSanchez:20} for obtaining cross sections for neutron-capture on unstable nuclei. Neutron-induced reactions play an important role in stellar nucleosynthesis and thus are critical to our understanding of the formation of neutron-rich nuclei heavier than iron. The experimental measurements of the reaction cross sections needed for such nuclear astrophysics simulations are difficult. The phenomenological approaches to inelastic scattering are not predictive as the nucleon-nucleus scattering potential is tuned to reproduce scattering observables in experiments. Transitions to each excited state observed in the experiment requires fitting a new parameter in scattering calculation. Therefore, it is critical to develop nuclear theory to reliably predict these cross sections and enable indirect measurements.  Such developments are indispensable. In the future even if the needed experiments become possible, to perform them for all reactions will remain a challenge. However if theoretical formalisms are available, one can use experiments as a benchmark to validate the theory and then apply it for the various cases needed. \\ 

In these proceedings we briefly discuss the two key components of our approach, the JLM folding model and the nuclear structure inputs we use to perform reaction calculations, and show some results for $^{90}$Zr, as a test-case for the machinery we are developing. Towards the end we analyze our results to identify areas of future development.\\

\section{Our Approach}
To study nucleon-nucleus scattering in a fully-microscopic picture, one needs to solve an $A+1$ body scattering problem using the interaction between the projectile and the $A$ nucleons of an $A$-body target. We approximate this $A+1$-body problem as a two-body quantum scattering problem, while explicitly taking the target's microscopic structure into account. This is done by combining the microscopic structure of the target with nucleon-induced inelastic scattering using a single-folding approach~\cite{Satchler:Book}.  In other words, the incoming unbound nucleon interacts with the correlated $A$ nucleons of the target in the nucleon-nucleus scattering process and can be described using the radial coupled channel equation in the center-of-mass frame of the nucleus and the target, 
\begin{equation} \label{CC}
\bigg{\{} \frac{d^2}{dr^2} - \frac{l_i(l_i+1)}{r^2} -\frac{2\mu}{\hbar^2} U_{ii} (r) + k_i^2\bigg{\}} \psi_i(r) = \sum_{f\neq i} \frac{2\mu}{\hbar^2} U_{if}(r) \psi_f(r).
\end{equation}
Here, $\mu$ is the reduced mass, $k_i$ is initial relative momentum of the projectile, and $U_{if}(r)$ are the nucleon-nucleus scattering potentials for a transition of the target from state $i$ to $f$. In single-folding, the effective two-body nucleon-nucleus interaction, $U_{if} (r)$ in Eq.~\ref{CC} is computed by folding target densities with a pre-selected parametrization of the $NN$ force between a projectile nucleon and the target nucleon. \\

In our work we use the effective $NN$ force (between projectile and target nucleon) first developed by Jeukenne, Lejeune and Mahaux (JLM)~\cite{JLM:74} and later improved by Bauge et al. (JLM-B)~\cite{Bauge:98,Bauge:01,Bauge:00}.  The scattering solutions are computed using FRESCO~\cite{thompson:88}, a publicly available coupled-channels code, in the Distorted-wave Born approximation (DWBA), i.e., we assume only one step target transitions from ground to excited states during direct inelastic scattering. 

\subsection{Jeukenne-Lejeune-Mahaux (JLM) folding model} 
The JLM folding model is based on an optical potential calculated in nuclear matter formed by free nucleons that are interacting via Reid's hard-core $NN$ force. The core idea is, at leading order a nucleus, as seen by the projectile, is approximated to be asymmetric nuclear matter. The potential energy density experienced by an incoming nucleon of energy $E$ in nuclear matter, consisting of proton(neutron) density $\rho_p(\rho_n)$, is parametrized as follows to match the on-shell nuclear matter g-matrix, 
\begin{align} \label{V_NN}
&V_{T=0}(\rho_p, \rho_n, E, V_c(r)) =  \lambda_V(E) \frac{V_0(\rho_0 ,E'(r))}{\rho_0} + \lambda_W(E) \frac{W_0(\rho_0 ,E'(r))}{\rho_0}  \\ \nonumber
&V_{T=1}(\rho_p, \rho_n, E, V_c(r)) = \tau\bigg{[}\lambda_V(E)\lambda_{V_1}(E) \frac{V_1(\rho_0,E'(r))}{\rho_0}  + \lambda_{W}(E)\lambda_{W_1}(E)\frac{W_1(\rho_0,E'(r))}{\rho_0}\bigg{]}
\end{align}
where $\rho_0 = \rho_p+\rho_n$, $E'(r) = E-V_c(r)$ and $\tau = 1 (-1)$ for incident proton (neutron). We note that $V_c(r)$ is the Coulomb potential experienced by the incoming projectile at radial distance $r$ from the center of the target. $T$ is the total isospin of the individual projectile and target nucleons. 
The central nucleon-nucleus scattering potentials and spin-orbit form factors are then calculated as 
\begin{align}\label{V_NA}
U_{T} (r,E, t_i, t_r) = (t\sqrt{\pi})^{-3} \int &\bigg{[}Re[V_{T}(\rho_p(r_t), \rho_n(r_t), E, V_c(r_t))] e^{-|\vec{r} - \vec{r_t}|^2/t_r^2}  \\ \nonumber
&+ Im[V_{T}(\rho_p(r_t), \rho_n(r_t), E, V_c(r_t))] e^{-|\vec{r} - \vec{r_t}|^2/t_i^2}\bigg{]}  \rho_T d^3\vec{r_t}, \\ \nonumber
U_{n(p)}^{\mathrm{SO}}(r) = & (\lambda_{v_{so}} + i\lambda_{w_{so}})\frac{1}{r} \frac{d}{dr} \bigg{(} \frac{2}{3}\rho_{p(n)} + \frac{1}{3} \rho_{n(p)}\bigg{)} 
\end{align}
where, $\rho_1= \rho_n - \rho_p$. Parameters $t_r$  and $t_i$ introduce finite range to the effective $NN$ interaction in a medium to account for variation in density of a finite nucleus. 
\subsubsection{Accounting for the density dependence of the effective interaction, the rearrangement term :} In inelastic scattering, the target nucleus undergoes an excitation due to which the target ground state density is perturbed. It is clear from the discussion above that the effective $NN$ interaction depends on the density of the medium. Therefore, one needs to include contributions from varying the in-medium $NN$ interaction as the target density changes, i.e., from $V_T^{(R)}(\rho_0,E, V_c(r)) = \frac{\partial V_T(\rho_0,E')}{\partial \rho_0}$. $V_T^{(R)}$ is also often called as ``the rearrangement term". The importance of including rearrangement while using JLM for inelastic scattering was first pointed out by Cheon et al.~\cite{Cheon85-1,Cheon85-2}. Later works~\cite{Dupuis:15, Dupuis:19} have shown that the effect from rearrangement strongly depends on the incident energy and the angular momentum transfer of the transition.  The inelastic scattering cross sections presented here include this term. In addition, for proton-nucleus scattering we also include the direct Coulomb term contribution to the nucleon-nucleus transition potential using a multipole expansion. As we discuss in upcoming sections, the implementation and impact of the Coulomb exchange and the spin-orbit terms on coupling potentials, however, remain an open question and will be investigated in future work. \\

The JLM and JLM-B parameterizations differ in the $\lambda$ values, the effective range parameters and the functional forms of the imaginary terms of the effective interaction in eqns.~(\ref{V_NN}-\ref{V_NA}). The calculations presented in these proceedings are performed using the JLM-B interaction, exact details of which are discussed in~\cite{Bauge:01}. The target ground state and transition densities used in above equations come from solving the $A$-body Schrödinger equation. \\ 

\subsection{Nuclear structure calculations as input}
We model the target as a many-nucleon system in which the nucleons interact with each other via the Gogny-D1M potential~\cite{Gogny:09}. The ground state of the target is calculated variationally in an axially-deformed basis using HFB many-body method~\cite{Younes:19, LLNLQRPA}. The HFB calculations provide the ground state energy and its mean-field density. Next, we build the vibrational excitation spectrum of the nucleus using QRPA~\cite{Peru:14}. Essentially the many-body excited states are calculated by finding eigenvalues and eigenfunctions for vibrational excitations of quasiparticles at the HFB minima. Using HFB and QRPA results, one can obtain the transition densities for excitations from ground to higher energy states. The scattering potentials in the folding approach use the HFB radial ground-state densities in the elastic channel, and the QRPA radial transition densities in the direct inelastic channels. We note that the use of axially-deformed harmonic oscillator as the computational basis will allow us to extend our approach to study scattering from deformed nuclei in the future. The details of the structure results for Zr isotopes used in these proceedings are available in~\cite{Chimanski:23, Chimanski:22}. \\
\label{sec-1}

\section{Scattering Results : $^{90}$Zr}
\begin{figure}[h!]%
    \centering
    \subfloat{{\includegraphics[width=6.1cm]{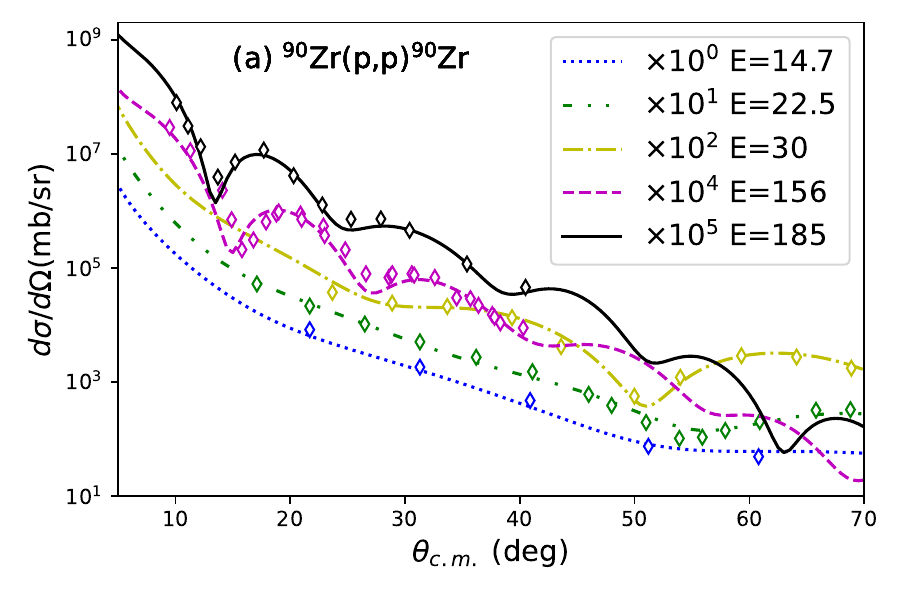} \label{fig:1a} }}%
    \qquad
    \subfloat{{\includegraphics[width=6.1cm]{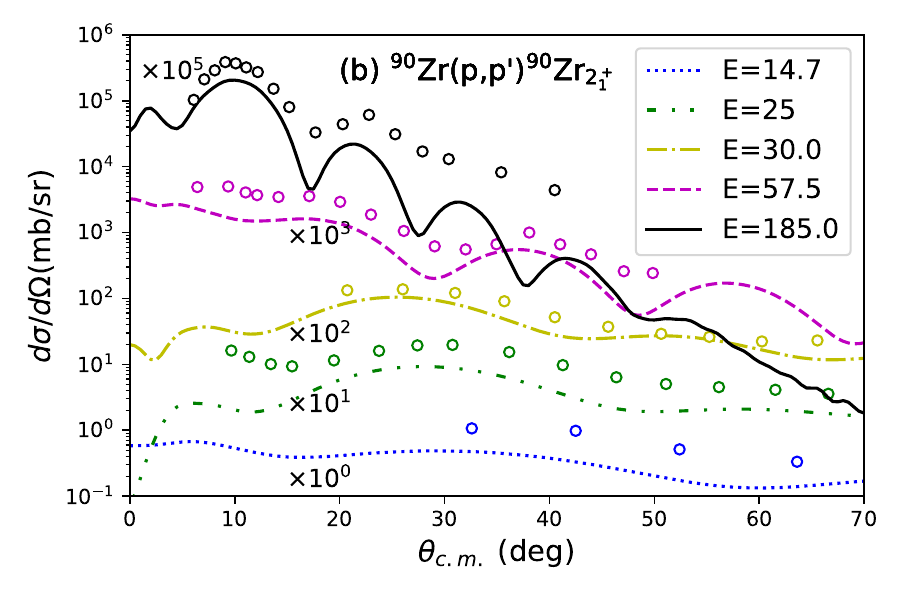}\label{fig:1b} }}%
    \subfloat{{\includegraphics[width=6.1cm]{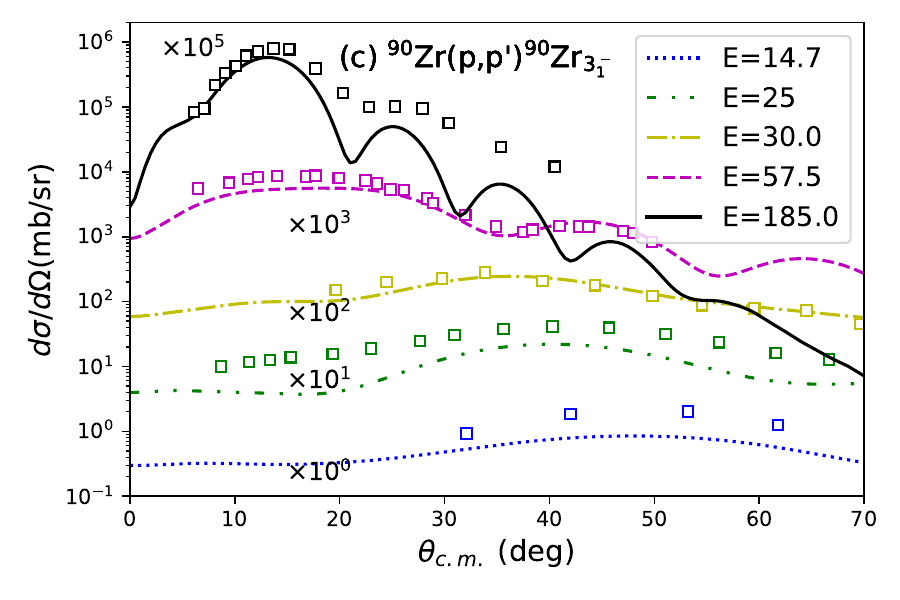}\label{fig:1c} }}%
    \caption{Comparison between differential cross sections from experimental data and from our approach for proton elastic (\ref{fig:1a}) and inelastic (\ref{fig:1b} and \ref{fig:1c}) scattering off of a $^{90}$Zr target. The curves are our results while the discrete markers are experimental values. Different proton beam energies (in MeV) are illustrated by different colors.}%
    \label{fig:1}%
\end{figure}

We choose $^{90}$Zr as the test case of the approach and our implementation. It is a well-studied stable spherical nucleus with sufficient experimental data for proton scattering to use as benchmarks. We begin with elastic angular differential cross section comparisons for the proton$-^{90}$Zr system, as shown in Figure~\ref{fig:1a}. The curves are our results in comparison to the diamonds that show experimental data for various incident proton energies, represented by different colors. Good agreement between our elastic scattering results and experimental data checks our implementation of the JLM folding approach and the scattering calculations. Next, we use the same volume terms of the $NN$ interaction, along with addition of the rearrangement term, to perform the inelastic scattering calculations using DWBA. \\

To perform proof-of-principle calculations for inelastic scattering, we consider the excitation of $^{90}$Zr to its first $2^{+}$ and $3^-$ excited states. Table~\ref{tab:1} compares the excitation energy, $E(J_1^\pi)$, and corresponding reduced electric quadrupole transition strength, $B(E_J : 0^+ \rightarrow J^\pi)$, from our HFB+QRPA calculation to accepted experimental data. Figure~\ref{fig:1b} and~\ref{fig:1c} show our results for the inelastic differential cross sections, indicated by curves, with respect to the available data that are represented by circles and squares for the target excitation to the first $2^+$ and $3^-$, respectively. The varying proton beam energies between 14.7 MeV and 185 MeV are shown by different colors. We find that in the case of inelastic scattering we consistently underpredict the differential cross sections for all incident energies. This trend is consistent with the lower electromagnetic transition strengths from the nuclear structure calculations. In addition, from table~\ref{tab:1} it is clear that the HFB+QRPA calculations reproduce the $3^-$ state better than the first $2^+$ excited state. This trend is reflected in the differential cross section results if one compares the degree of agreement between the theoretical and the experimental cross sections for the inelastic scattering to the first $2^+$ with the first $3^-$, at a fixed beam energy. This demonstrates the impact of microscopic nuclear structure on scattering predictions. \\

It is important to note that we use the same structure input for all incident energies. However, from figures~\ref{fig:1b} and~\ref{fig:1c} it is clear that the agreement of the theoretical results with experimental data depends on the beam energy. The results presented in these proceedingss do not include the Coulomb exchange and the spin-orbit contributions to the transition potentials. Our initial investigations suggest that the contribution of the Coulomb exchange term to the transition potentials impacts the differential cross sections, especially at lower energies. The importance of including the full Coulomb contribution to the coupling potential for low-lying excitations is known in the context of another single-folding potential~\cite{khoa:02}. As the incident beam energy increases, our preliminary analysis also shows that the spin-orbit contribution to the coupling potentials becomes increasingly important, especially at higher scattering angles. 

\begin{table}
\centering
\caption{$^{90}$Zr : HFB+QRPA nuclear structure results~\cite{Chimanski:23}, used in the folding approach, in comparison to experimental data for the first 2$^{+}$ and 3$^{-}$ excited states.}
\label{tab:1} 
\begin{tabular}{lll}
\hline
 Observable & Theory & Exp.  \\\hline
$E(2_1^+)$ [MeV]&2.725 & 2.186 \\
$B(E_2 : 0^+ \rightarrow 2_1^+) [e^2 b^2]$ &0.038 & 0.061~\cite{BE2Tab:16}\\\hline
$E(3_1^-)$ [MeV]&2.858& 2.748 \\ 
$B(E_3 : 0^+ \rightarrow 3_1^-) [e^2 b^3]$ &0.078 & 0.098~\cite{Kibedi:02} \\\hline
\end{tabular}
\end{table}

\section{Conclusions and Outlook}
We implemented the JLM folding model approach in combination with state-of-the-art nuclear structure methods, HFB and QRPA with the Gogny-D1M potential, to compute scattering observables. The central idea is that integrating the structure and reaction theory will enable us to predict the outcome of scattering experiments. In this way, the structure of the nucleus informs by design the reaction calculations about the changes the reaction causes in that target. The nuclear structure (HFB+QRPA) codes developed at LLNL~\cite{LLNLQRPA} were carefully checked against CEA Bruyere implementation~\cite{Peru:14}. Similarly, the folding procedures at LLNL~\cite{LLNLFC} and the CEA~\cite{Dupuis:19} were compared to each other. Finally, both laboratories use different reaction codes~\cite{thompson:88, ECIS}. Such careful testing provides confidence in the overall machinery and allows us to focus on physics developments. We compare our results for proton scattering on $^{90}$Zr to data. Our implementation of the JLM-B parametrization of the JLM potential, used to fold the HFB ground-state density, reproduces the direct elastic scattering cross sections in the energy range of 14.7 MeV-185 MeV, consistent with the validity range of 1 keV-200 MeV for the JLM-B nucleon-nucleus optical potential. \\ 

In the case of inelastic scattering, however, we find that inelastic differential cross sections are underpredicted, and the agreement between theoretical results and experimental data varies with changing beam energy. We show that better agreement between the theoretical and experimental excitation energies and the reduced electromagnetic transition strengths implies better reproduction of scattering cross sections. This was demonstrated by comparing our results for the first $2^+$ and $3^-$ excited states. This is in agreement with previous work that uses JLM for inelastic scattering off of $^{208}$Pb and $^{209}$Bi nuclei~\cite{Dupuis:19}. The energy dependence of the disagreement between our scattering results and the experimental data is likely a consequence of missing spin-orbit and Coulomb exchange term contributions to the folded transition potentials. The spin-orbit term is proportional to the angular-momentum transfer between target and projectile. Therefore, one would expect that the importance of the spin-orbit component grows with increasing beam energy as higher partial waves become accessible. In contrast, at energies close to the Coulomb barrier, and at very forward angles where the long range part of the potential plays a dominant role, it is important to include the full Coulomb potential that can excite the target. Hence at lower beam energies, we expect improvement in our results from including the Coulomb exchange contribution to the transition potential. \\

Our next step is to develop the prescription to include the spin-orbit and Coulomb exchange terms in the coupling potential for the JLM approach. To reliably use our approach to predict scattering observables for unstable neutron-rich nuclei, it is important that we improve it for stable nuclei where experimental data is readily available. 

%
%
%

\begin{thebibliography}{26}

\bibitem{Ratkiewicz:19prl}
A.~Ratkiewicz, J.A. Cizewski, J.E. Escher, G.~Potel, J.T. Burke, R.J.
  Casperson, M.~McCleskey, R.A.E. Austin, S.~Burcher, R.O. Hughes et~al., Phys.
  Rev. Lett. \textbf{122}, 052502 (2019)

\bibitem{Escher:18prl}
J.E. Escher, J.T. Burke, R.O. Hughes, N.D. Scielzo, R.J. Casperson, S.~Ota,
  H.I. Park, A.~Saastamoinen, T.J. Ross, Phys. Rev. Lett. \textbf{121}, 052501
  (2018)

\bibitem{Escher:16a}
{J. E. Escher}, {A. P. Tonchev}, {J. T. Burke}, {P. Bedrossian},
  {R. J. Casperson}, {N. Cooper}, {R. O. Hughes}, {P. Humby}, {R.
  S. Ilieva}, {S. Ota} et~al., EPJ Web of Conferences \textbf{122}, 12001 (2016)

\bibitem{Escher:12rmp}
J.E. Escher, J.T. Burke, F.S. Dietrich, N.D. Scielzo, I.J. Thompson, W.~Younes,
  Rev. Mod. Phys. \textbf{84}, 353 (2012)

\bibitem{PerezSanchez:20}
R.~P\'erez~S\'anchez, B.~Jurado, V.~M\'eot, O.~Roig, M.~Dupuis, O.~Bouland,
  D.~Denis-Petit, P.~Marini, L.~Mathieu, I.~Tsekhanovich et~al., Phys. Rev.
  Lett. \textbf{125}, 122502 (2020)

\bibitem{Satchler:Book}
G.R. Satchler, \emph{Direct nuclear reactions} (Oxford University Press, 1983)

\bibitem{JLM:74}
J.P. Jeukenne, A.~Lejeune, C.~Mahaux, Phys. Rev. C \textbf{10}, 1391 (1974)

\bibitem{Bauge:98}
E.~Bauge, J.P. Delaroche, M.~Girod, Phys. Rev. C \textbf{58}, 1118 (1998)

\bibitem{Bauge:01}
E.~Bauge, J.P. Delaroche, M.~Girod, Phys. Rev. C \textbf{63}, 024607 (2001)

\bibitem{Bauge:00}
E.~Bauge, J.P. Delaroche, M.~Girod, G.~Haouat, J.~Lachkar, Y.~Patin, J.~Sigaud,
  J.~Chardine, Phys. Rev. C \textbf{61}, 034306 (2000)

\bibitem{thompson:88}
I.J. Thompson, Computer Physics Reports \textbf{7}, 167 (1988)

\bibitem{Cheon85-1}
T.~Cheon, K.~Takayanagi, K.~Yazaki, Nuclear Physics A \textbf{437}, 301 (1985)

\bibitem{Cheon85-2}
T.~Cheon, K.~Takayanagi, K.~Yazaki, Nuclear Physics A \textbf{445}, 227 (1985)

\bibitem{Dupuis:15}
{M. Dupuis}, {E. Bauge}, EPJ Web of Conferences \textbf{122}, 06001 (2016)

\bibitem{Dupuis:19}
M.~Dupuis, G.~Haouat, J.P. Delaroche, E.~Bauge, J.~Lachkar, Phys. Rev. C
  \textbf{100}, 044607 (2019)

\bibitem{Gogny:09}
S.~Goriely, S.~Hilaire, M.~Girod, S.~P\'eru, Phys. Rev. Lett. \textbf{102},
  242501 (2009)

\bibitem{Younes:19}
W.~Younes, D.~Gogny, J.F. Berger, \emph{A Microscopic Theory of Fission
  Dynamics Based on the Generator Coordinate Method} (Springer, Cham, 2019)

\bibitem{LLNLQRPA}
W.~Younes, \emph{Work in progress (2023)}

\bibitem{Peru:14}
S.~{P{\'e}ru}, M.~{Martini}, European Physical Journal A \textbf{50}, 88 (2014)

\bibitem{Chimanski:23}
E.V. Chimanski, E.J. In, J.E. Escher, S.~Péru, W.~Younes (2023),
  \texttt{arXiv:2308.13374}

\bibitem{Chimanski:22}
E.V. Chimanski, E.J. In, J.E. Escher, S.~Péru, W.~Younes, Journal of Physics:
  Conference Series \textbf{2340}, 012033 (2022)

\bibitem{khoa:02}
D.~Khoa, E.~Khan, G.~Colò, N.~{Van Giai}, Nuclear Physics A \textbf{706}, 61
  (2002)

\bibitem{BE2Tab:16}
B.~Pritychenko, M.~Birch, B.~Singh, M.~Horoi, Atomic Data and Nuclear Data
  Tables \textbf{107}, 1 (2016)

\bibitem{Kibedi:02}
T.~Kibédi, R.~Spear, Atomic Data and Nuclear Data Tables \textbf{80}, 35
  (2002)

\bibitem{LLNLFC}
A.~Thapa, J.~Escher, \emph{Work in progress (2023)}

\bibitem{ECIS}
J.~Raynal, \emph{Computer code ECIS06, NEA 0850/19},
  \urlstyle{tt}\url{https://www.oecd-nea.org/tools/abstract/detail/nea-0850}

\end{thebibliography}

\section*{Acknowledgement}
This work was performed under the auspices of the U.S. Department of Energy by Lawrence Livermore National Laboratory under Contract DE-AC52-107NA27344, with partial support from LDRD projects 20-ERD-030, 19-ERD-017. The work at Brookhaven National Laboratory was sponsored by the Office of Nuclear Physics, Office of Science of the U.S. Department of Energy Contract No. DE-AC02-98CH10886 with Brookhaven Science Associates, LLC.

\end{document}